\documentclass[conference,letter]{IEEEtran}
\ifCLASSINFOpdf
\else

\fi

\usepackage{graphicx}
\usepackage{epstopdf}
\usepackage{balance}
\usepackage{multicol}
\usepackage{amsmath}
\usepackage{mathtools}
\usepackage{amssymb}
\usepackage{esint}
\usepackage{amsfonts}
\usepackage{multirow}
\usepackage{wrapfig}
\usepackage{cite}
\usepackage{subcaption}
\usepackage{lipsum}
\usepackage{float}    
\usepackage{algorithm}
\usepackage{caption}
\usepackage[none]{hyphenat}
\usepackage[usenames,dvipsnames]{color}
\usepackage[letterpaper, left=0.7in, right=0.7in, bottom=0.75in, top=0.75in]{geometry}

\begin{document}
\title{\LARGE Federated Mimic Learning for Privacy Preserving Intrusion Detection}
\author{\IEEEauthorblockN{Noor Ali Al-Athba Al-Marri, Bekir S. Ciftler,
and Mohamed M. Abdallah}
\IEEEauthorblockA{Division of Information and Computing Technology, College of Science and Engineering,\\
Hamad Bin Khalifa University, Doha, Qatar\\
naalmarri@mail.hbku.edu.qa, \{bsciftler, moabdallah\}@hbku.edu.qa}\vspace{-9mm}}

\maketitle
\begin{abstract}
Internet of things (IoT) devices are prone to attacks due to the limitation of their privacy and security components.
These attacks vary from exploiting backdoors to disrupting the communication network of the devices.
Intrusion Detection Systems (IDS) play an essential role in ensuring information privacy and security of IoT devices against these attacks.
Recently, deep learning-based IDS techniques are becoming more prominent due to their high classification accuracy.
However, conventional deep learning techniques jeopardize user privacy due to the transfer of user data to a centralized server.
Federated learning (FL) is a popular privacy-preserving decentralized learning method.
FL enables training models locally at the edge devices and transferring local models to a centralized server instead of transferring sensitive data.
Nevertheless, FL can suffer from reverse engineering ML attacks that can learn information about the user's data from model.
To overcome the problem of reverse engineering, mimic learning is another way to preserve the privacy of ML-based IDS.
In mimic learning, a student model is trained with the public dataset, which is labeled with the teacher model that is trained by sensitive user data.
In this work, we propose a novel approach that combines
the advantages of FL and mimic learning, namely federated mimic learning to create a distributed IDS while minimizing the risk of jeopardizing users' privacy, and benchmark its performance compared to other ML-based IDS techniques using NSL-KDD dataset.
Our results show that we can achieve 98.11\% detection accuracy with federated mimic learning.
\end{abstract}
\section{Introduction}
\label{sect:Introduction}
Smart and interconnected internet of things (IoT) devices became prominent in our daily life as they provide users with vital services through human-to-machine or machine-to-machine communications \cite{Mosenia2017Comprehensive}.
Smart devices in smart homes help users to turn on the heater, lock the doors, and monitor through cameras and smart alarms to ensure the safety and security of home.
Due to the high level of connectivity risks of IoTs, adversaries can generate intrusive attacks on IoT devices to control or to monitor the user behavior\cite{Bugeja2017analysis}.
Therefore, safeguarding IoT devices against these attacks is vital.

Researchers proposed various solutions to prevent intrusion attacks, which rely on analyzing the network.
These solutions are called Intrusion Detection Systems (IDS) in general.
Recently, with the growth in data availability and processing power, machine learning (ML) based techniques became prominent in IDS. 
However, ML-based approaches jeopardize the privacy of the end users\cite{liu2017smart}.
The privacy of users endangered because existing ML-based techniques rely on transferring the user data, which contains sensitive information of the user's behavior to a centralized server for processing and training ML models.
Hence, preserving user's data privacy is becoming the focus of the ML-based IDS research.

Federated learning (FL) is a revolutionary distributed machine learning technique that utilizes the computational power of edge devices\cite{mcmahan2016communication} without exchanging data samples of users.
The local models are trained with user data at the device, and these models are transmitted to the centralized server.
Hence, FL is partially privacy-preserving and communication-wise efficient since it avoids the transmission of vast amounts of data\cite{park2019wireless,niknam2019federated}.
However, the privacy preservation of FL may suffer from reverse engineering since it is trained on sensitive user data.

Recently, as an alternative solution to the privacy problem, mimic learning is proposed by \cite{Shafee2020Mimic} to preserve the privacy of end-users.
The research focuses on using mimic learning in order to allow data transmission of intrusion detection knowledge from a teacher model to a student model.
The authors indicate that the performance of both teacher and student model is nearly identical, although the datasets for both of them are different.
It proves that the unlabeled dataset that is trained by the teacher model can be used in order to transfer knowledge to the student model without disclosing any sensitive information.
Nevertheless, mimic learning provides a trained model from a single user; hence it is not distributed and may not provide a broad solution.

In this paper, we propose a novel solution that integrates FL, along with mimic learning for developing an ML-based IDS.
Utilizing the advantages of both techniques, we create a privacy-preserving distributed ML-based IDS.
We utilize the NSL-KDD dataset to benchmark our proposed system against existing solutions.
In our system model, we preprocess the dataset as well as applying feature selection to reduce the computational load on the edge devices.
Our machine learning model is based on MultiLayer Perceptron (MLP), and the same model with the same parameters were used in all scenarios for a fair benchmark. To the best of our knowledge, we are the first to apply federated mimic learning for privacy-preserving purposes which will help in providing up-to-date intrusion detection systems.

Our main contributions with this study are listed below;
\begin{itemize}
\item We have developed a novel federated mimic learning-based IDS technique by taking advantage of both FL and mimic learning to minimize the possibility of obtaining any sensitive data against reverse engineering attacks on the student model. 
\item We have implemented the system using Python on Google Colab and carried out simulations using the real-world dataset (NSL-KDD) to benchmark our proposed model against centralized and federated ML-based IDS. 
\end{itemize}

This paper is structured as follows.
In Section~\ref{sect:RelatedWork}, a brief overview of existing literature on IDS is presented.
We explain our proposed method of federated mimic learning for the IDS in Section~\ref{sect:SystemModel}.
The dataset and the preprocessing techniques used in our simulations are explained in Section~\ref{sect:Preprocessing}.
Subsequently, we analyze the numerical results for the performance benchmark of our proposed method in Section~\ref{sect:Results}.
Finally, we present our concluding remarks in Section~\ref{sect:Conclusion}.
\section{Literature Review}
\label{sect:RelatedWork}
Machine learning techniques are becoming pre-eminent in the design of IDS due to their capability of extracting hidden connections in the data.
In this section, we will review several ML-based IDS in the literature as well as the studies about the available datasets for IDS.
In \cite{sikder20176thsense}, the authors proposed '6th Sense' which is an IDS that helps in strengthening IoT security by detecting changes found in sensor data.
It then creates a relative model to differentiate between malicious and non-malicious behaviors of target sensors.
The proposed system utilizes three types of ML-based detection mechanisms: Markov Chain, Naive Bayes, and Logistic Model Tree.
The proposed solution collects sensor data of user activity by custom made applications.
The final result of the data analysis shows if the current state of the device is considered as malicious or not.
Finally, the proposed solution achieved accuracy over 95\% using the three ML methods, which makes it highly effective and efficient when detecting sensor-based attacks.
    

As a solution to privacy problems, the authors in \cite{meidan2019privacy} investigate how to ensure user's behavioral information and traffic data privacy when implementing an ML-based IDS behind a network address translation (NAT) in a smart home.
The local detector, between the router and the optical network terminal, monitors the network address translated (NATed) traffic data emerging from the home network, and the pre-trained classifier is applied to detect IoT devices.
Results show that using the proposed solution allows service providers to detect 73\% of any NATed IoT device.
However, this approach requires manual updating of the classifiers.

A study by \cite{rawat2019intrusion} uses the NSL-KDD data set to analyze it using ML techniques as well as deep learning algorithms to implement and evaluate IDS in the networks.
A deep neural network was trained, and the following scenarios were considered to compare the models using NSL-KDD:
\begin{itemize}
    \item Classifying the records of network connections either as a normal connection or an attack according to the features existing in the NSL-KDD dataset.
    \item Classifying the records of network connections either as a normal connection or an attack according to the minimum number of features in the NSL-KDD dataset. 
\end{itemize}


A study by \cite{nkiama2016subset} proposed a feature selection mechanism that focuses on excluding non-relevant features to classify features that will contribute to the improvement of the detection rate, based on the performance of each feature during the selection process.
A recursive feature elimination procedure was used to classify the correct related features and associate them with a decision tree-based classifier.
    
A survey paper by \cite{sherasiya2016survey} analyzed the different types of existing IDS such as signature-based IDS which is based on analyzing the network traffic and compare the signature against predefined attack patterns, anomaly-based IDS that is based on understanding what is considered normal behavior in the network and in case it detects behavior that is different from usual then it defines it as an intrusion attack, as such it is considered to be more efficient compared to the signature-based in terms of detection.
Lastly is specification-based IDS, which defines the normal behavior that is done manually, therefore ensures the reduction of false-positive rate.


In \cite{Mourabit2014Wireless}, the authors proposed an anomaly detection-based approach that suggested a mobile agent-based intrusion detection program within Wireless Sensor Networks (WSN).
It uses a multi-agent and classification-based intrusion detection.
The proposed system has fewer parameters to characterize the attacks so that work can be enhanced by creating more complex detection parameters and using statistical detection of anomalies to enable the creation of signatures for the attack. 

The authors of \cite{Karuppiah2014Novel} proposed a hierarchical energy-efficient IDS to detect Sybil node in WSN.
The system proposed lays down two cases: In the first case, a centralized approach is developed to send an acknowledge queried data packets.
The cluster head manages and maintains a table that is used to store names and locations of all nodes.
In the second case, all valid nodes with their identities and current position coordinates respond to the cluster head.
Additionally, Sybil node sends its identities and current position, so that cluster head matches such data in a table with valid nodes. 
If any problems occurred, the Sybil node is identified.
The outcome of the simulation reveals that the proposed system increases energy efficiency and reliably detects the Sybil node.

In \cite{Shafee2020Mimic}, mimic learning is utilized to deploy privacy-preserving ML-based IDS.
The teacher model is trained on private labeled data with four different types of classifiers: Decision Tree Induction (DTI), Random Forest (RF), Support Vector Machine (SVM), and Naïve Bayes (NB).
The classifier with the highest result is then selected, and the teacher model is used to label an unlabeled public dataset.
The newly labeled public dataset is used to train the same four types of classifiers to generate the student models for each.
The student model is used as a privacy-preserving knowledge transfer.
Results show that the RF classifier for both the teacher and student model had the highest accuracy in detection while NB was the lowest.


\begin{figure}[t]
    \centering
    \includegraphics[width=0.95\linewidth]{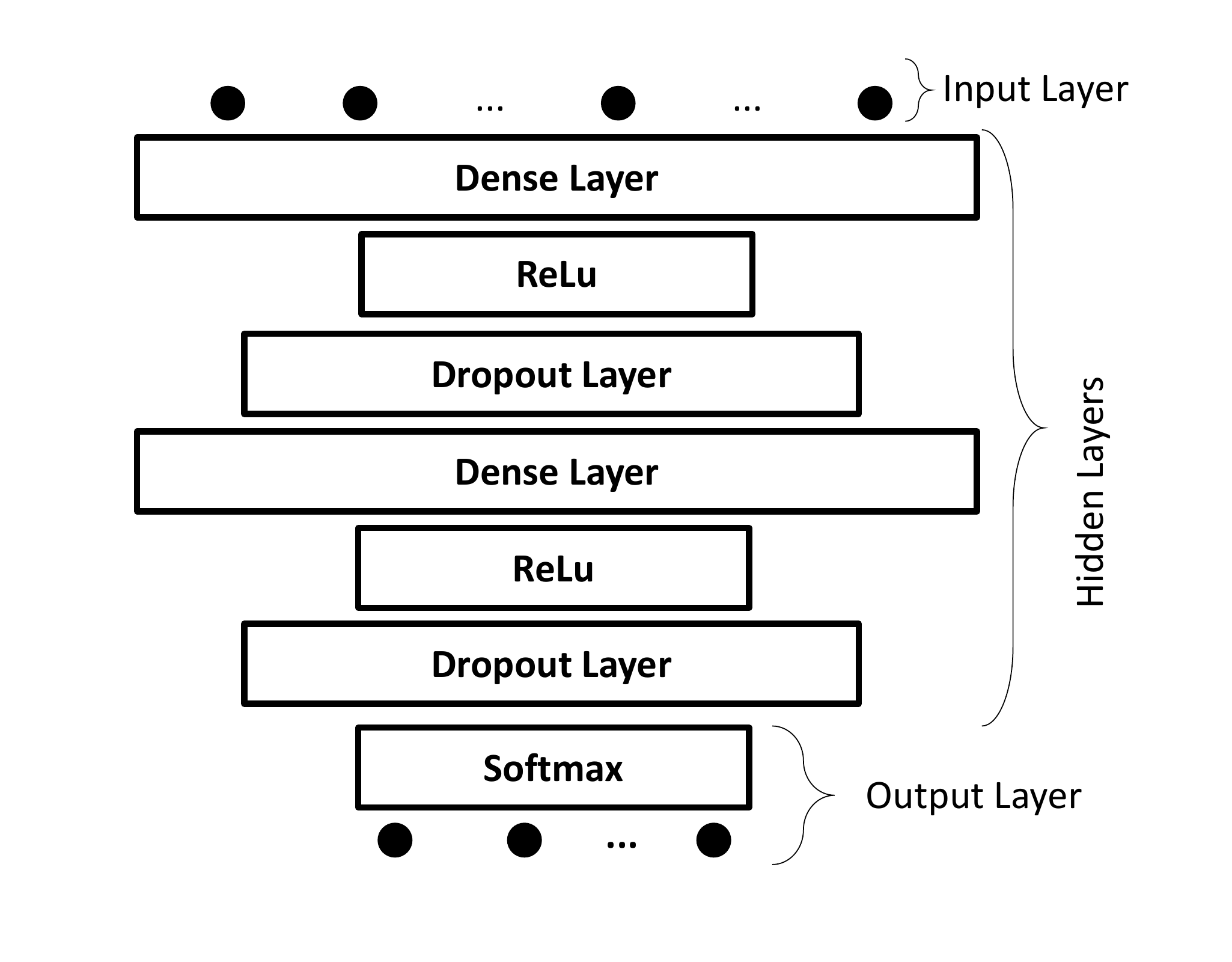}
    \caption{Feedforward Multilayer Perceptron Architecture.}
    \label{fig:MLP}
\end{figure}

\begin{figure}[t]
    \centering
    \begin{subfigure}[t]{0.23\textwidth}
        \centering
        \includegraphics[width=0.95\linewidth]{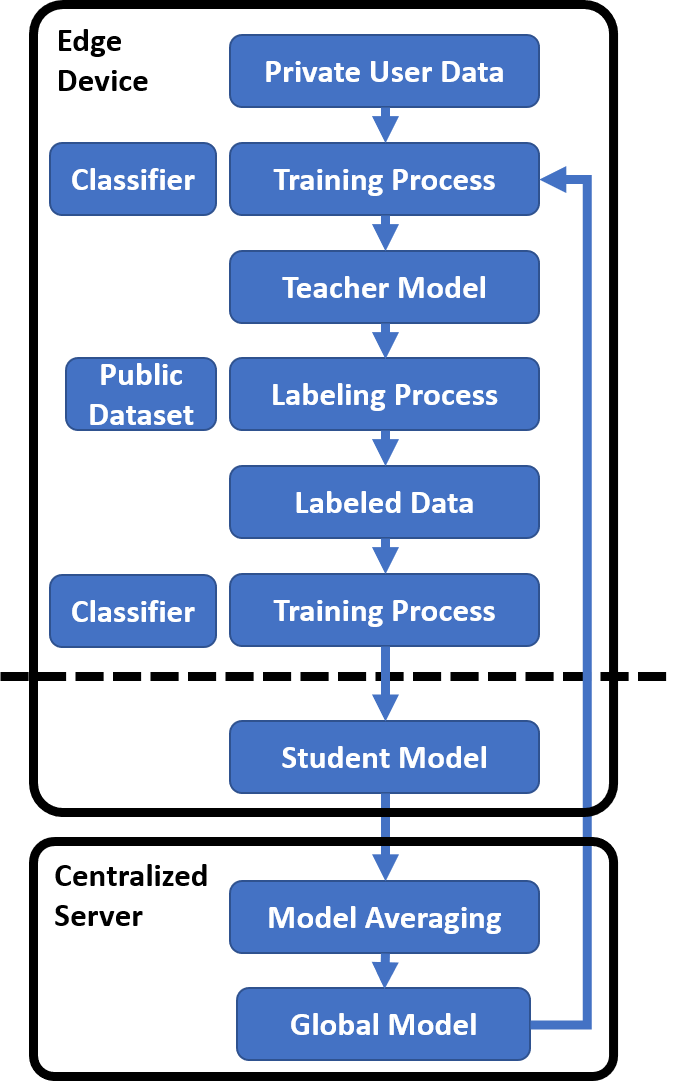}
        \caption{FTML}
    \end{subfigure}~\begin{subfigure}[t]{0.23\textwidth}
        \centering
        \includegraphics[width=0.95\linewidth]{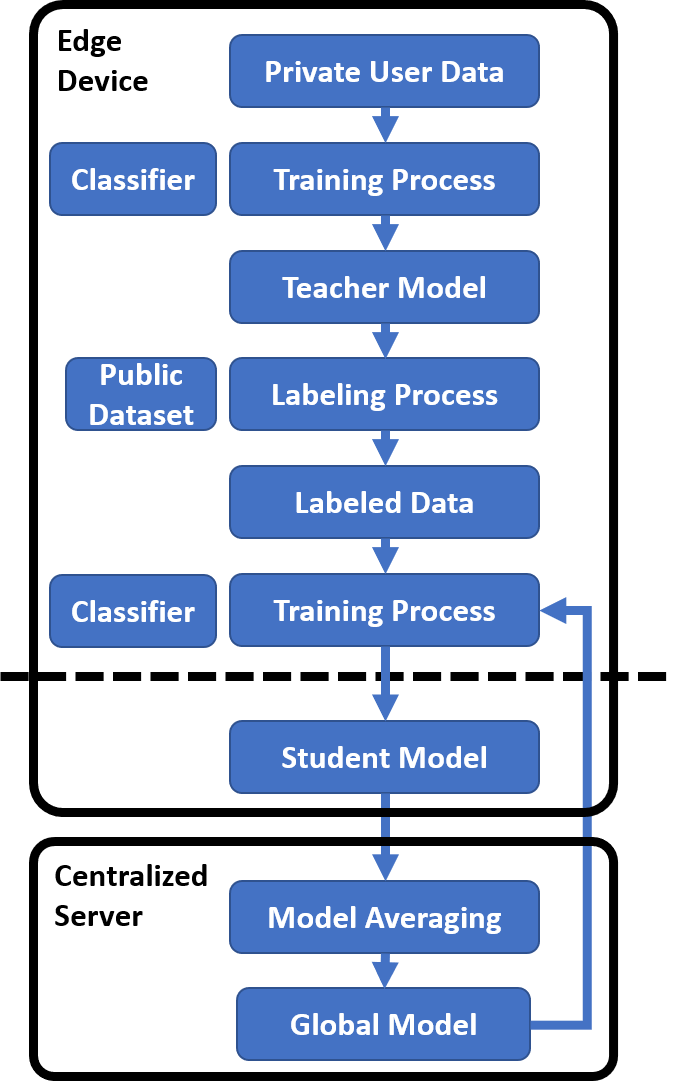}
        \caption{FSML}    
    \end{subfigure}
    \caption{System Models for Federated Teacher Mimic Learning (FTML) and Federated Student Mimic Learning (FSML).}
    \label{fig:SystemModel}
    \vspace{-5mm}
\end{figure}
\section{Federated Mimic Learning for Privacy-Preserving Intrusion Detection}
\label{sect:SystemModel}

Feedforward Multilayer Perceptrons (MLPs) are used as our neural network architecture for the IDS, as shown in Fig.~\ref{fig:MLP}.
We created two hidden layers with 256 neural units each.
For activation function, we use the Rectified Linear Unit (ReLU).
We use a dropout rate of 0.4 to ensure regularization after every hidden layer.
Dropout layers help in controlling over-fitting by removing an individual unit with a random probability while training the model.
The softmax activation function is used for the output layer of the classifier.

\subsection{Federated Mimic Learning}

The implementation of federated mimic learning is based on FL, as shown in Fig.\ref{fig:SystemModel}.
The simulation and learning parameters are kept the same with Deep Learning and FL for a fair benchmark.
The teacher models of users are created utilizing each users' private training dataset.
The teacher models are utilized to label the unlabeled public dataset at each user.
The labeled public dataset of each user is then used to generate the student models.
Then the student models of each user are transferred to the centralized server for federated averaging to create the new global model.
There are two methods for implementing federated mimic learning.
If the averaged global model is returned to the user and included in the loop of training from the teacher model, it is called Federated Teacher Mimic Learning (FTML).
If the global model is included in the loop of training from the student model, it is called Federated Student Mimic Learning (FSML).
The pseudo-codes of FTML and FSML are shared with Algorithm~1 and Algorithm~2.

\begin{figure}[h]
    \centering
    \includegraphics[width=0.5\textwidth]{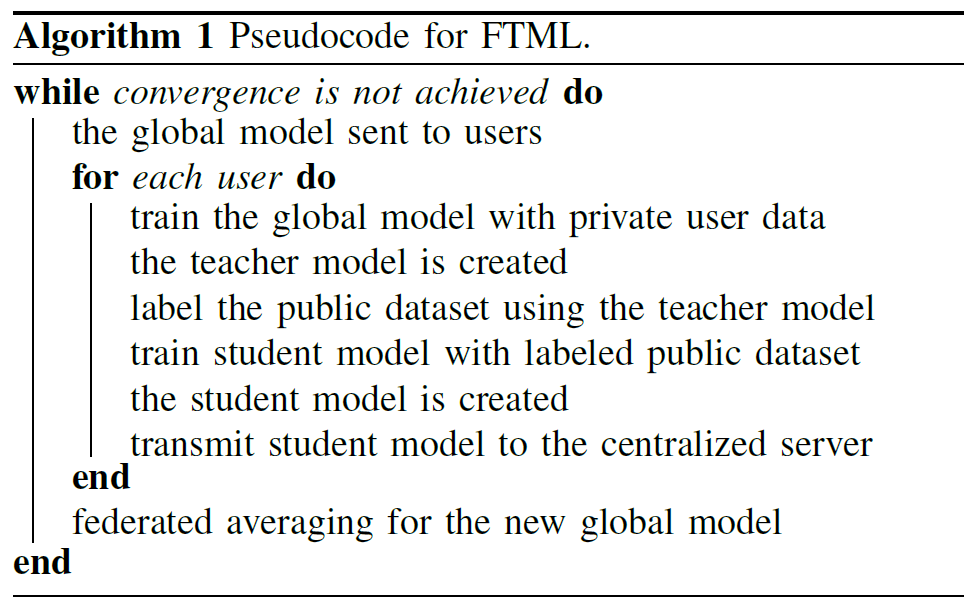}
\end{figure}

\begin{figure}[h]
    \centering
    \includegraphics[width=0.5\textwidth]{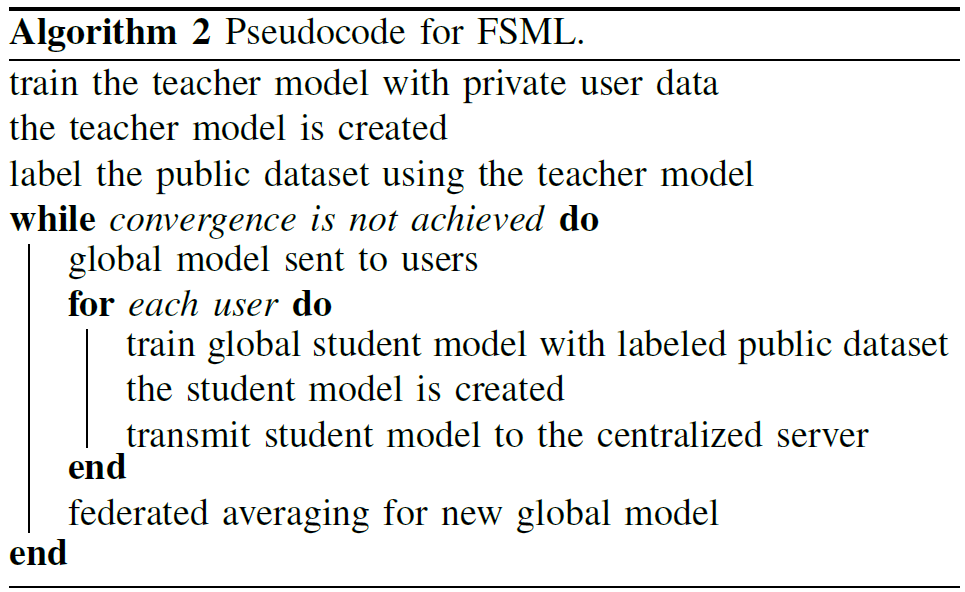}
\end{figure}

\section{Dataset and Preprocessing}
\label{sect:Preprocessing}

\subsection{NSL-KDD Dataset}

\begin{table}[H]
\centering
\caption{Number of samples in training set and test set.}
\begin{tabular}{|l|c|c|}
\hline
\textbf{Label} & \multicolumn{1}{l|}{\textbf{\# of Training Samples}} & \multicolumn{1}{l|}{\textbf{\# of Test Samples}} \\ \hline
\textbf{(0) DoS} & 41,334 & 4,592 \\ \hline
\textbf{(1) Normal} & 60,608 & 6,734 \\ \hline
\textbf{(2) Probe}  & 10,490 & 1,165 \\ \hline
\textbf{(3) R2L}    & 895    & 99    \\ \hline
\textbf{(4) U2R}    & 46     & 5     \\ \hline
\end{tabular}
\label{tab:Samples}
\end{table}

NSL-KDD dataset is one of the most used traffic data for developing IDS.
The number of samples for the dataset for each type are summarized in Table~\ref{tab:Samples}. 
It is an enhanced version of the KDD'99 dataset, which had some drawbacks in terms of the massive amount of repeated records and dataset imbalance, which resulted in easily detecting attack classes.
The NSL-KDD dataset was introduced to overcome such drawbacks.
This dataset includes 41 features which belong to 3 major families:
\begin{itemize}
\item Basic Features: Features associated with connection information such as hosts, ports, protocols, and services used. 
\item Traffic Features: Features that are calculated as a collection during window interval.
\item Content Features: Features that are obtained from data packet or payload and are related to a certain protocol or application used.
\end{itemize}

Every row (sample) in the NSL-KDD dataset contains a label to indicate whether it is normal connectivity or a particular type of attack.
The dataset includes four distinctive attack classes:
\begin{itemize}
    \item Denial of Service (DOS): A cyber-attack that targets a device or a machine to make it unavailable to the intended users either temporarily or indefinitely by disrupting its services.
    \item Probe: The initial step of an attack. The attacker gathers information on web applications, operating systems, databases, networks, and the devices connecting to it. Attackers scan them to identify both known and unknown vulnerabilities.
    \item User to Root (U2R): An attack that allows an attacker to gain root privileges when accessing a machine.
    \item Remote to Local (R2L): An unauthorized access from a remote device to a local device.
\end{itemize}
We mapped each attack label to five classes: four are the attacks mentioned above or normal.
After mapping them, we added an output column called "Attack" that indicates the type of the attack or normal connectivity.

\subsection{One-Hot Encoding for Categorical Data}
There are three kinds of features in the NSL-KDD dataset: nominal, binary, and numeric. 
Binary data are data that contains numeric values, which is enough to indicate their presence by either (0) or (1).
Nominal data are variables that include categorical values instead of numeric values.
Deep learning techniques cannot operate on such data; therefore, one-hot encoding is applied to nominal features in order to change them to numeric features.

Additionally, since we also apply one-hot encoding, it will transform into 70 new features while the "flag" feature will transform into 11 new features. Therefore, the original 41 features will become 127 features in the dataset. 

After transforming all the nominal features into numeric using one-hot encoding, we apply normalization in order to range the values between $0$ to $1$.
This allows us to balance the dataset from having vast numbers that might affect our model accuracy due to imbalanced classifiers.

\subsection{Feature Elimination}

Feature selection with logistic regression helps to identify the critical features in a dataset.
It is a crucial step when tuning a model.
It helps in reducing the dimensionality of the dataset, which enhances the performance and speed of a model.
This is followed by recursive feature elimination (RFE) which selects smaller sets of features of the dataset.
Critical features can be obtained using feature importance attribute in RFE. 
In order to select the critical features for each class of attack and normal class, we implemented recursive feature elimination with logistic regression in order to get the top 20 features for detecting each type of attack.
We were able to differentiate each type of attack as well as the normal cases by setting 1 and 0 as the output of training samples.
This process is repeated in the dataset until the desired number of selected features is obtained.
Out of 127 features, 42 features were selected as a result of logistic regression-based feature selection. 
We then used these selected features to feed it into our model for both training and testing.

\begin{table}[h]
    \centering
    \caption{System parameters for the simulations.}
    \begin{tabular}{cc}
        \hline
        \multicolumn{1}{|l|}{\textbf{Parameter}} & \multicolumn{1}{l|}{\textbf{Value}} \\ \hline\hline
        \multicolumn{1}{|l|}{Deep Learning Libraries} & \multicolumn{1}{l|}{Google TensorFlow \& Keras} \\ \hline
        \multicolumn{1}{|l|}{Optimizer} & \multicolumn{1}{l|}{Adam} \\ \hline
        \multicolumn{1}{|l|}{Learning Rate} & \multicolumn{1}{l|}{$0.001$} \\ \hline
        \multicolumn{1}{|l|}{$\beta_1$, $\beta_2$} & \multicolumn{1}{l|}{$0.1$, $0.99$} \\ \hline
        \multicolumn{1}{|l|}{Number of Hidden Layers} & \multicolumn{1}{l|}{$2$} \\ \hline
        \multicolumn{1}{|l|}{Number of Hidden Nodes} & \multicolumn{1}{l|}{$256$ nodes in each hidden layer} \\ \hline
        \multicolumn{1}{|l|}{Activation Function} & \multicolumn{1}{l|}{ReLu} \\ \hline
        \multicolumn{1}{|l|}{Loss Function} & \multicolumn{1}{l|}{Mean Absolute Error} \\ \hline
        \multicolumn{1}{|l|}{{Batch Size}} & \multicolumn{1}{l|}{{$128$}}  \\ \hline
        \multicolumn{1}{|l|}{{Number of Epochs}} & \multicolumn{1}{l|}{{$10$}}  \\ \hline
        \multicolumn{1}{|l|}{{Number of Rounds}} & \multicolumn{1}{l|}{{$20$}}  \\ \hline
        \label{table:parameter}
    \end{tabular}
    \vspace{-5mm}
\end{table}

\section{Numerical Results}
\label{sect:Results}

The preprocessed NSL-KDD dataset was used as the input for our deep learning model.
We apply a cross-validation method to evaluate our model performance.
The dataset is split into $90\%$ for the training set and $10\%$ for the testing set. Therefore, the training dataset $113,375$ samples, while for the testing set, it is $12,598$.
The ADAM optimizer is used for training the model, while the batch size is $128$, and the learning rate is set to $0.001$.
Our model and learning algorithms are implemented using TensorFlow based architecture on the Python environment.
Code evaluations were implemented on Google Colab with TPU acceleration.
The simulations parameters are shown in Table~\ref{table:parameter}.

\begin{table}[h]
\caption{DL-based classifier accuracy results.}
\begin{tabular}{|l|c|c|c|c|c|}
\hline
\textbf{Label} &
  \multicolumn{1}{l|}{\textbf{Accuracy}} &
  \multicolumn{1}{l|}{\textbf{Precision}} &
  \multicolumn{1}{l|}{\textbf{Recall}} &
  \multicolumn{1}{l|}{\textbf{FalseAlarm}} &
  \multicolumn{1}{l|}{\textbf{F-Score}} \\ \hline
\textbf{Normal} & 98.6  & 98  & 100 & 2.86 & 99 \\ \hline
\textbf{DoS}    & 98.89 & 99  & 98  & 0.54 & 98 \\ \hline
\textbf{Probe}  & 99.8  & 100 & 98  & 0.03 & 99 \\ \hline
\textbf{R2L}    & 99.26 & 0   & 0   & 0    & 0  \\ \hline
\textbf{U2R}    & 99.95 & 0   & 0   & 0    & 0  \\ \hline
\end{tabular}
\vspace{-5mm}
\label{tab:deepL}
\end{table}
\begin{figure}[h]
    \centering
    \includegraphics[width=0.3\textwidth]{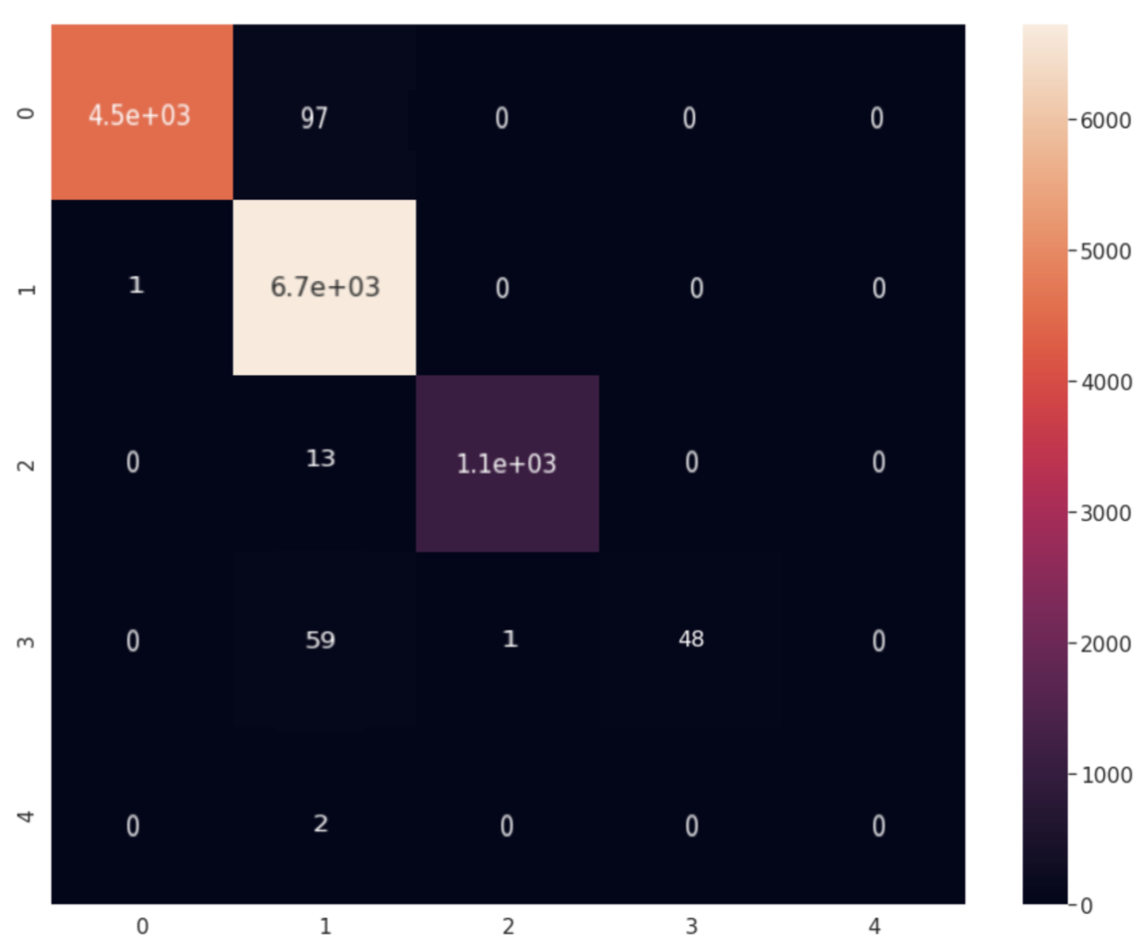}
    \caption{Confusion matrix for DL-based classifier.}
    \label{fig:deepFig}
\end{figure}

The result indicates that our deep learning model is capable of detecting attacks that are generated on IoT devices traffic with a percentage equal to 98.15\% for the training, while the test accuracy is equal to 98.28\%.
Results of centralized deep learning-based IDS is summarized in Table~\ref{tab:deepL} and Fig.~\ref{fig:deepFig} shows the confusion matrix for centralized deep learning-based classifier.

As a second step, the FL-based solution is implemented to benchmark its performance against centralized deep learning.
In this case, the user's data is not transmitted to the centralized intrusion detection provider.
Instead, 5000 data points are distributed over ten users and used to train local models for each user.
Then, each user provides their trained local models to the centralized server of the intrusion detection system.
The centralized server applies model averaging to local models to create a new global model.
The global model is returned to individual users for federated training.
This loop is considered a round.
After 20 rounds,
the results indicate that with preserving user's privacy, we were able to achieve a 98.61\% detection accuracy using FL.
This result shows that privacy of users can be enhanced by utilizing edge resources with high accuracy. 
Table~\ref{tab:federatedL} and Fig.~\ref{fig:federatedFig} below summarizes the result of each attack category using the confusion matrix in FL-based classifier.
\begin{table}[h]
\caption{FL-based classifier accuracy results.}
\begin{tabular}{|l|c|c|c|c|c|}
\hline
\textbf{Label} &
  \multicolumn{1}{l|}{\textbf{Accuracy}} &
  \multicolumn{1}{l|}{\textbf{Precision}} &
  \multicolumn{1}{l|}{\textbf{Recall}} &
  \multicolumn{1}{l|}{\textbf{FalseAlarm}} &
  \multicolumn{1}{l|}{\textbf{F-Score}} \\ \hline
\textbf{Normal} & 98.62 & 98  & 100  & 2.93  & 99   \\ \hline
\textbf{DoS}    & 99.13 & 100 & 98   & 0.006 & 99   \\ \hline
\textbf{Probe}  & 99.9  & 100 & 99   & 0.02  & 100  \\ \hline
\textbf{R2L}    & 99.59 & 100 & 0.45 & 0     & 0.62 \\ \hline
\textbf{U2R}    & 99.95 & 0   & 0    & 0     & 0    \\ \hline
\end{tabular}
\label{tab:federatedL}
\vspace{-5mm}
\end{table}
\begin{figure}[h]
    \centering
    \includegraphics[width=0.3\textwidth]{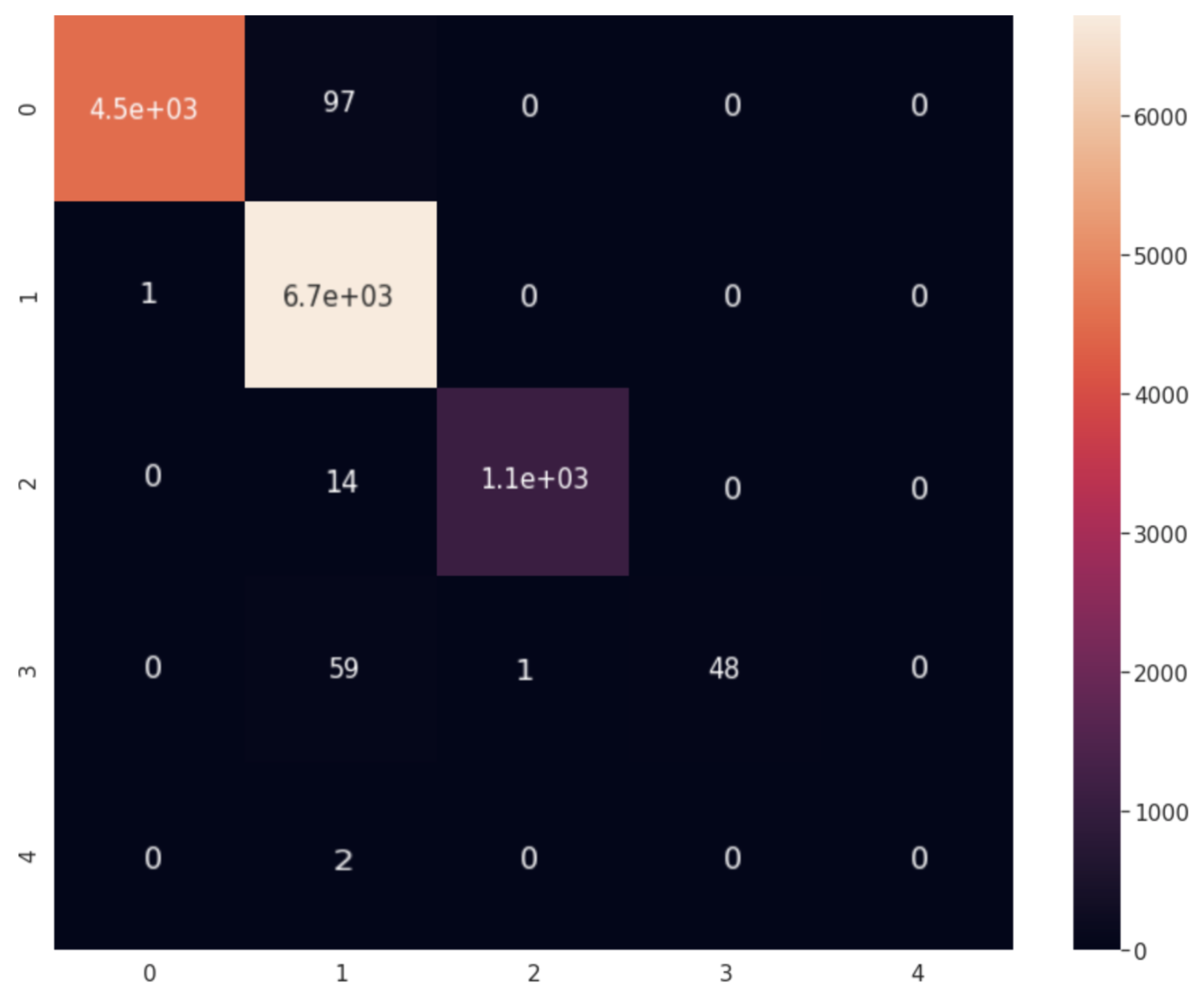}
    \caption{Confusion matrix for FL-based classifier.}
    \label{fig:federatedFig}
\end{figure}

To further enhance data privacy, we have implemented federated mimic learning, as explained in Section~\ref{sect:SystemModel}. To achieve that, we divided the training data like the following: 60\% of the dataset is used in the private dataset for teacher models, while the remaining 40\% were used as the unlabeled public dataset for student models.
The teacher models are trained at ten users, and these models are used to label the unlabeled public dataset at each user.
Each user trains its student model with the public dataset labeled with its teacher model.
The student models are then averaged to obtain the global model.
The global model returned to the training of either teacher model as in FTML or student model as in FSML.
The functionality of the proposed method was tested using the test dataset. Results indicate that with the use of the proposed federated mimic learning-based method, we were able to achieve a 98.118\% detection accuracy using FTML.
Table~\ref{tab:my-table1} and Fig.~\ref{fig:my_label1} below summarize the accuracy and the confusion matrix for FTML, respectively.
\begin{table}[h]
\caption{FTML-based classifier accuracy results.}
\begin{tabular}{|l|c|c|c|c|c|}
\hline
\textbf{Label} &
  \multicolumn{1}{l|}{\textbf{Accuracy}} &
  \multicolumn{1}{l|}{\textbf{Precision}} &
  \multicolumn{1}{l|}{\textbf{Recall}} &
  \multicolumn{1}{l|}{\textbf{FalseAlarm}} &
  \multicolumn{1}{l|}{\textbf{F-Score}} \\ \hline
\textbf{Normal} & 98.14 & 97  & 100 & 3.93 & 98 \\ \hline
\textbf{DoS}    & 99.09 & 100 & 98  & 0.03 & 99 \\ \hline
\textbf{Probe}  & 99.87 & 100 & 99  & 0.02 & 99 \\ \hline
\textbf{R2L}    & 99.14 & 0   & 0   & 0    & 0  \\ \hline
\textbf{U2R}    & 99.98 & 0   & 0   & 0    & 0  \\ \hline
\end{tabular}
\label{tab:my-table1}
\end{table}
\begin{figure}[h]
    \centering
    \includegraphics[width=0.3\textwidth]{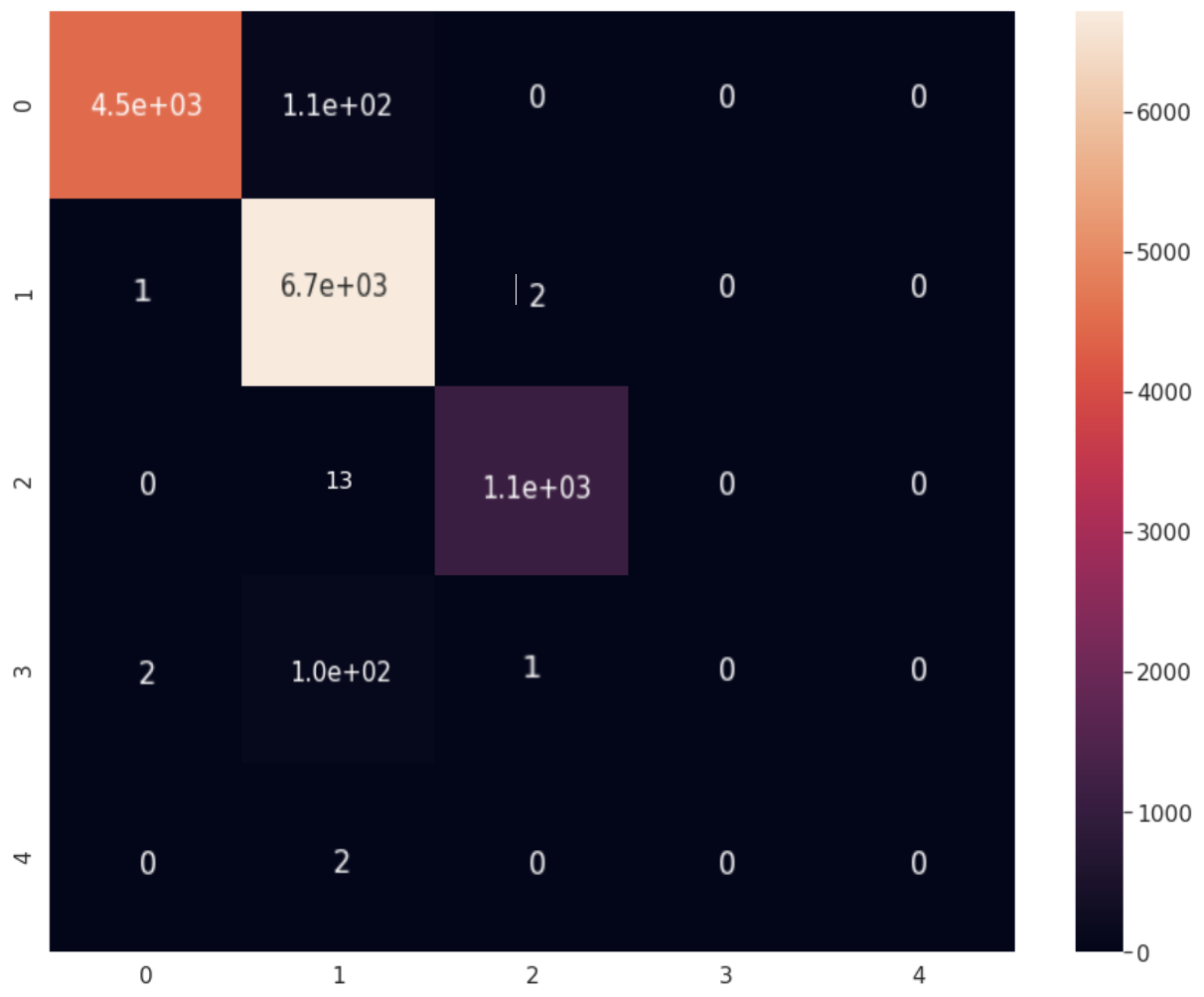}
    \caption{Confusion matrix for FTML-based classifier.}
    \label{fig:my_label1}
    \vspace{-3mm}
\end{figure}

In addition to that, we tested the detection accuracy in FSML.
Results indicate that with the use of federated mimic learning global model in the training process of the student model, we were able to achieve a 98.110\% detection accuracy using federated student mimic learning.
Table~\ref{tab:my-table2} and Fig.~\ref{fig:my_label2} below summarizes the result with respect to each attack category and the confusion matrix for FSML.
\begin{table}[h]
\caption{FSML-based classifier accuracy results.}
\begin{tabular}{|l|c|c|c|c|c|}
\hline
\textbf{Label} &
  \multicolumn{1}{l|}{\textbf{Accuracy}} &
  \multicolumn{1}{l|}{\textbf{Precision}} &
  \multicolumn{1}{l|}{\textbf{Recall}} &
  \multicolumn{1}{l|}{\textbf{FalseAlarm}} &
  \multicolumn{1}{l|}{\textbf{F-Score}} \\ \hline
\textbf{Normal} & 98.13 & 97  & 100 & 3.95 & 98 \\ \hline
\textbf{DoS}    & 99.08 & 100 & 98  & 0.03 & 99 \\ \hline
\textbf{Probe}  & 99.87 & 100 & 99  & 0.02 & 99 \\ \hline
\textbf{R2L}    & 99.14 & 0   & 0   & 0    & 0  \\ \hline
\textbf{U2R}    & 99.98 & 0   & 0   & 0    & 0  \\ \hline
\end{tabular}
\label{tab:my-table2}
\vspace{-1mm}
\end{table}
\begin{figure}[h]
    \centering
    \includegraphics[width=0.3\textwidth]{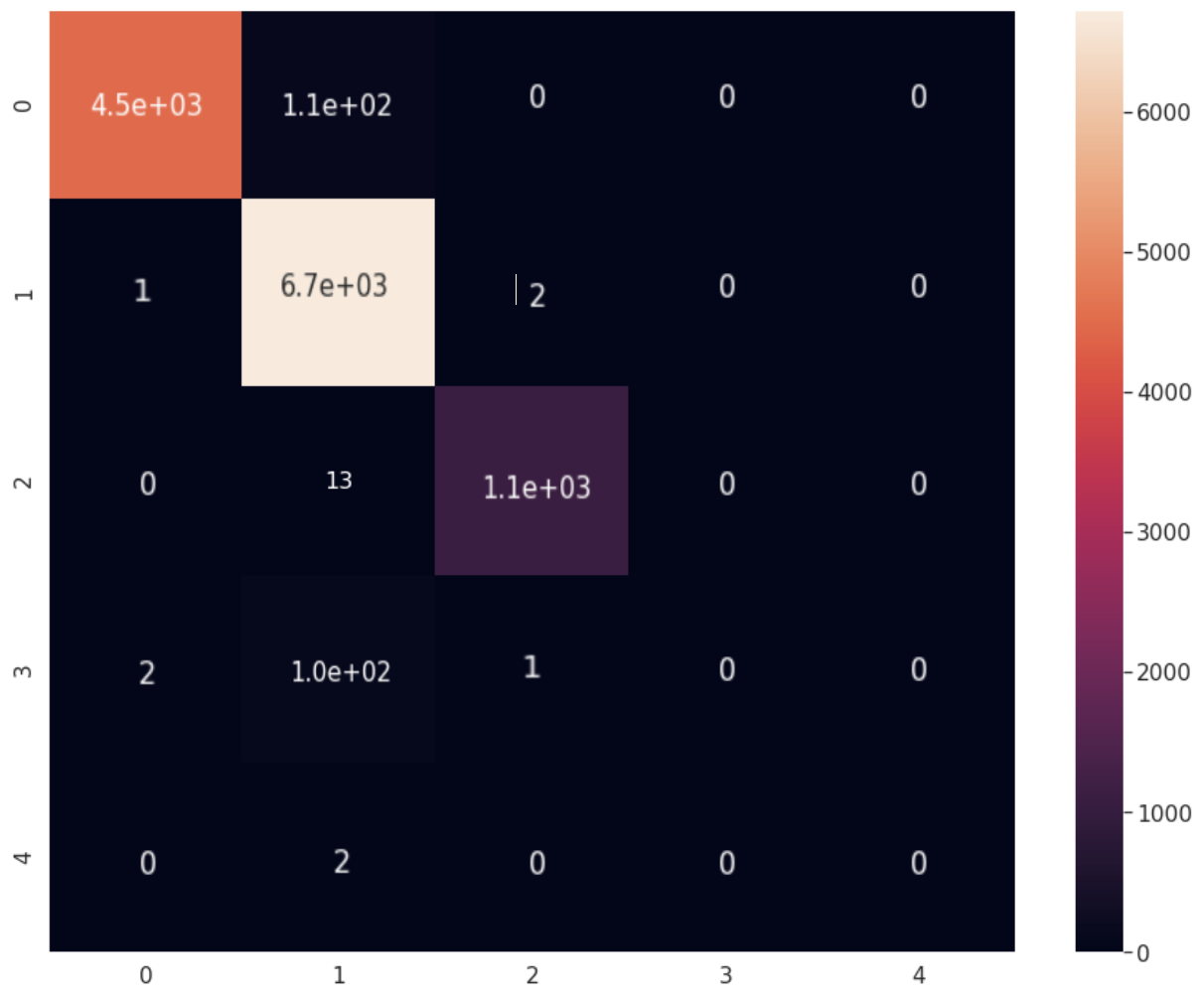}
    \caption{Confusion matrix for FSML-based classifier.}
    \label{fig:my_label2}
\end{figure}

Note that FSML requires half the computational cost of FTML same as since it only requires the device to train local model once per round, instead of two as in FTML.
As a result, FSML can achieve close performance to the centralized deep learning and FTML, while improving the privacy preservation of the user data significantly.

\begin{table}[]
\centering
\begin{tabular}{|l|l|c|}
\hline
\textbf{Classification Algorithm} & \textbf{Class Name} & \multicolumn{1}{l|}{\textbf{Accuracy}} \\ \hline
\multirow{5}{*}{\textbf{Random Forest}} & Normal & 99.1  \\ \cline{2-3} 
                                        & DoS    & 98.7  \\ \cline{2-3} 
                                        & Probe  & 97.6  \\ \cline{2-3} 
                                        & R2L    & 96.8  \\ \cline{2-3} 
                                        & U2R    & 97.5  \\ \hline
\multirow{5}{*}{\textbf{Naive Bayes}}   & Normal & 70.3  \\ \cline{2-3} 
                                        & DoS    & 72.7  \\ \cline{2-3} 
                                        & Probe  & 70.9  \\ \cline{2-3} 
                                        & R2L    & 69.8  \\ \cline{2-3} 
                                        & U2R    & 70.7  \\ \hline
\multirow{5}{*}{\textbf{FTML}}          & Normal & 98.14 \\ \cline{2-3} 
                                        & DoS    & 99.09 \\ \cline{2-3} 
                                        & Probe  & 99.87 \\ \cline{2-3} 
                                        & R2L    & 99.14 \\ \cline{2-3} 
                                        & U2R    & 99.98 \\ \hline
\multirow{5}{*}{\textbf{FSML}}          & Normal & 98.13 \\ \cline{2-3} 
                                        & DoS    & 99.08 \\ \cline{2-3} 
                                        & Probe  & 99.87 \\ \cline{2-3} 
                                        & R2L    & 99.14 \\ \cline{2-3} 
                                        & U2R    & 99.98 \\ \hline
\end{tabular}
\caption{Results of proposed FTML and FSML model compared to Random Forest and Naive Bayes\cite{Revathi2013ADA}.}
\label{tab:CompareResults}
\end{table}

A study by \cite{Revathi2013ADA} analyzes various ML techniques and their detection performance for IDS using the NSL-KDD dataset. According to the authors, Random Forest (RF) classifier achieved a 99.1\% detection accuracy for normal traffic. Furthermore, RF achieved 98.7\% detection accuracy for DoS attacks, while its detection accuracy for Probe is 97.6\%. R2L and U2R detection accuracy using RF is 96.8\% and 97.5\%. Additionally, Naive Bayes achieved 70.3\% detection accuracy for detecting normal traffic. As for the attack classes, Naive Bayes achieved 72.7\% detection accuracy for DoS attacks, 70.9\% for Probe, 69.8\% for R2L, and 70.7\% for U2R. By comparing the above with our results, our proposed model shows that both FTML and FSML have higher detection accuracy compared to RF. Similarly, Naive Bayes has the lowest detection accuracy with a huge difference compared to our proposed model, as shown in Table~\ref{tab:CompareResults}. 
 
\section{Conclusion}
\label{sect:Conclusion}
In this study, we propose an ML-based method of IDS for IoT devices using federated mimic learning for preserving user privacy.
The paper was divided into three types of implementation: first, we implemented a centralized deep learning model of the IDS, then  is implemented.
After that, we implemented the proposed federated mimic learning method covering both federated teacher mimic learning, and federated student mimic learning as an ML-based IDS.
Results show that federated mimic learning provides a detection accuracy while maintaining privacy similar to deep learning and models by 98.118\% in federated teacher mimic learning (FTML).
Additionally, we obtained a result with a minimal difference while optimizing the half the computational cost of FTML with the federated student mimic learning (FSML) with a 98.11\% detection accuracy. To the best of our knowledge, we are the first to apply federated mimic learning for privacy-preserving purposes, which will help in providing up-to-date intrusion detection systems.

\balance
\bibliographystyle{IEEEtran}
\bibliography{refs}
\end{document}